\documentclass[prb,twocolumn, superscriptaddress]{revtex4}
\usepackage{graphicx,amsmath,amssymb,amsxtra}
\renewcommand{\narrowtext}{\begin{multicols}{2} \global\columnwidth20.5pc}

\def\be{\begin{eqnarray}}
\def\ee{\end{eqnarray}}

\newcommand{\Eq}[1]{Eq.~(\ref{#1})}

\newcommand{\Fig}[1]{Fig.~(\ref{#1})}

\newcommand{\cC}{{\cal C}}
\newcommand{\cH}{{\cal H}}

\newcommand{\cL}{{\cal L}}

\newcommand{\ket}[1]{{|#1\rangle}}

\begin{document}
\title{
Linear independence of nearest-neighbor valence bond states in several two-dimensional lattices
}
\author{Julia Wildeboer$^1$ and Alexander Seidel}
\affiliation{
Department of Physics and Center for Materials Innovation, 
Washington University, St. Louis, MO 63136, USA}
\date{\today}
\begin{abstract}
We show for several two-dimensional lattices that the nearest neighbor valence bond states are linearly independent. 
To do so, we utilize and generalize a method that was recently introduced and applied to the kagome lattice by one of the authors.  
This method relies on the choice of an appropriate cell for the respective lattice, for which a certain local 
linear independence property can be demonstrated. Whenever this is achieved, linear independence follows 
for arbitrarily large lattices that can be covered by such cells, for {\em both} open and periodic boundary conditions. We report that 
this method is applicable to the kagome, honeycomb, square, squagome, two types of pentagonal, square-octagon, 
the star lattice, two types of archimedean lattices, three types of  ``martini'' lattices, and to fullerene-type lattices, e.g., the 
well known ``Buckyball''.  
Applications of the linear independence property, such as the derivation of  effective quantum dimer models, 
or the constructions of new solvable spin-$1/2$ models, are discussed.
\end{abstract}
\maketitle
\section{Introduction} 
Quantum Heisenberg models and their extensions
are prime examples of simple toy models that provide realistic descriptions of
complicated emergent phenomena in interacting many-particle systems.
Under most circumstances, these models describe systems that
order magnetically at low temperatures,  in general agreement with 
the experimental situation. 
There has been much interest, however, in mechanisms 
leading to ground states that remain magnetically
disordered even at the lowest temperatures. 
Various scenarios exist for such a possibility, where we focus
on the important special case of systems with spin-$1/2$ degrees of freedom on a lattice.
In a valence bond crystal, the ground state is adiabatically
connected
to one where lattice spins are paired up into singlets, or ``valence bonds''.
Depending on the lattice, this may be possible with or
without\cite{shastrysutherland} the breaking of a
spatial symmetry. Other variants of singlet ``crystal''
phases feature ``singlet plaquettes'' instead of
individual valence bonds.
Even more interesting, however, is the case where adiabatic 
continuity to a trivial product state does not exist,
and the zero temperature spin state is devoid of any crystalline
character but forms a ``spin liquid'' driven by fluctuations of
valence bonds.
 This possibility was first considered by 
Anderson in 1973,\cite{anderson1}
and was coined a ``resonating valence bond'' (RVB) spin 
liquid. 

\begin{figure}[tb]
\centering
\includegraphics[width=3.6cm]{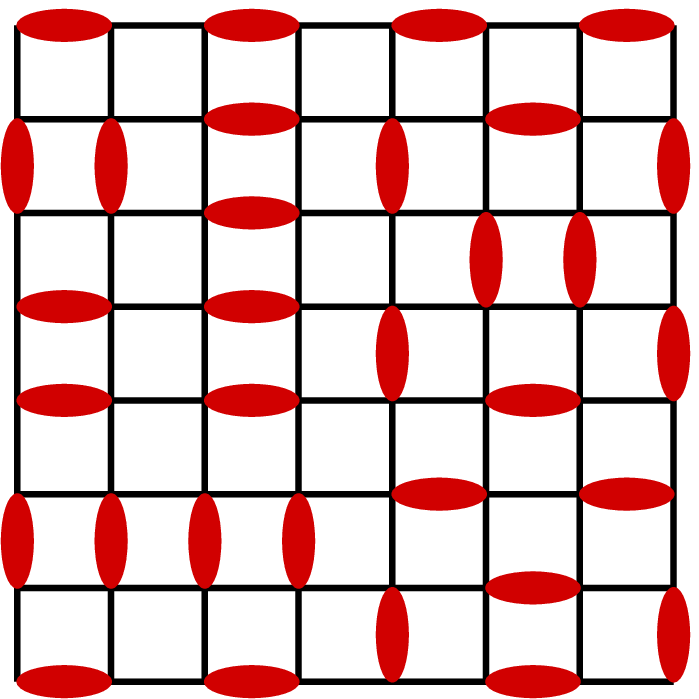}
\caption{
(Color online.) A square lattice with dimer covering.
Dimers are indicated by ovals.\label{dimers}
 }
\end{figure}

Interest in RVB spin liquid physics has been driven
both by its proposed connection\cite{anderson2}
to high $T_c$ superconductivity,
and by the innate exotic character of RVB states, which
feature fractionalized spin-$1/2$ excitations.
Promising experimental candidates have been identified only 
recently.\cite{triangular1, triangular2, kagome1, kagome2, kagome3, kagome4}
Theoretical challenges in establishing the existence of
RVB spin liquids have been profound, due to
the strongly interacting nature in particular of
$SU(2)$-invariant quantum spin systems.
To render the problem tractable, Rokhsar and Kivelson
invented an ingenious scheme to explore 
the non-magnetic part of the
phase diagram of quantum spin-$1/2$ systems through effective
``quantum dimer'' models (QDMs).\cite{RK}
They focused on the case where a gap
in the system renders all correlations short ranged.
In this case, the RVB spin liquid ground state
can be thought of as superposition of states
where spins pair up into short range valence bonds.
A quantum dimer model is obtained by first
truncating the Hilbert space
to include only states where each spin participates
in a nearest neighbor valence bond (NNVB).
The second simplification, perhaps even bolder
and more difficult to control, is to regard the NNVB states that generate the Hilbert
space as an {\em orthogonal} basis. In reality,
no two NNVB states on a finite lattice are orthogonal.
It is thus more appropriate to think of the degrees of freedom
of these new effective theories not as valence bonds, but as
hardcore bosons or ``dimers'' living on the
links of the original lattice. As sets, however,
both the hard core dimer states and the NNVB states are in one-to-one
correspondence with dimerizations of the lattice into
nearest neighbor pairs, see Fig. \ref{dimers}.

The exploration of QDMs has given rise to
profound insights into possible realizations of short range RVB
spin liquid physics, in particular on non-bipartite lattices.\cite{moessner,misguich}
It has remained challenging, however, to 
rigorously
establish the status of simple QDMs
as viable effective theories for quantum spin-$1/2$ systems within a certain parameter regime.
The lack of orthogonality of the NNVB states
that QDMs seek to describe makes it difficult to
establish a direct mapping between QDMs and 
the low energy sector of quantum spin-$1/2$ models.
This difficulty can be dealt with by treating the non-orthogonality
as a ``small parameter'', and setting up a systematic expansion
in this parameter. This notion already played a central role
in the original literature,\cite{RK, bsutherland} 
and was recently
explored in great detail in a series of insightful papers.\cite{poilblanc,schwandt}
Within this scheme, one can thus get the issue of the non-orthogonality of the NNVB states
under control. However, the validity of this perturbative scheme depends
crucially on the fact that the NNVB states, while not orthogonal,
are at least linearly independent, like their counterparts in QDMs.
In technical terms, the overlap matrix obtained from the NNVB states
must be invertible. The need for 
an invertible overlap matrix
was noticed early on,\cite{RK} 
and from thereon linear independence of NNVB states was routinely quoted as an
assumption in the literature, e.g. in estimates of the
low temperature entropy of highly frustrated quantum magnets.\cite{elser, nussinov}
Furthermore, exactly solvable, $SU(2)$-invariant spin-$1/2$ models 
with RVB and/or spin liquid ground states on simple lattices have only
been constructed quite recently,\cite{fujimoto,a_seidel09, cano, yao}
in addition to work on decorated lattices.\cite{decorate}
In Ref. \onlinecite{a_seidel09}, rigorous (albeit partial) statements
on the uniqueness of the RVB-type ground states of the model
constructed there were intimately tied to the
linear independence of NNVB states on the kagome lattice.
We also note that from a purist point of view,
there is a need to demonstrate that 
superpositions of NNVB wave functions,
which may be considered as variational\cite{bsutherland, elser, nussinov, klein, noorbakhsh}
or exact\cite{fujimoto,a_seidel09, cano} solutions
to various problems,
do not vanish
identically, whenever the overlaps
between the NNVB states forming these wave functions 
do not have a uniform sign.
The normalizability of such wave functions is an obvious byproduct of the
linear independence of NNVB states (on the
respective lattice).
The explicit or implicit assumption of
the linear independence of the NNVB states
 is thus a prevalent theme
in the literature on short range RVB physics, and in some 
cases has been studied extensively on finite clusters.\cite{mambrini, ML2}

Rigorous proofs of this linear independence have been available since 1989,
through a seminal work of Chayes, Chayes, and Kivelson.\cite{kcc}
The proof, however, has been limited to three different types of planar lattices,
the square, honeycomb, and square-octagon lattice,
and only for the case of open boundary conditions.
Here we discuss a more general method, that can, in principle,
be applied to any lattice, in the presence of both open 
and periodic boundary conditions. 
While we usually have Born--von Karman 
periodic boundary conditions in mind which
give a rectangular (or parallelogram) lattice strip
the topology of a torus, 
our method applies to other lattice topologies
as well. To demonstrate this, we
also apply our method to the C$_{60}$ lattice
and other fulleren-type lattices,
where the linear independence of NNVB
(or ``Kekul\'e'') states has direct applications
in chemistry.\cite{klein}

Although there is no guarantee
that our proof strategy works for every lattice where the linear independence
holds, we demonstrate its applicability to many new two-dimensional (2D) lattices,
for which the linear independence of NNVB states is first
established in this work. At the same time, we generalize
the aforementioned previous results on linear independence
of NNVB states
to the case of periodic boundary conditions.
It is well known that the physics of short range RVB states
becomes enriched in subtle ways when periodic boundary 
conditions are imposed. On a toroidal
square lattice, e.g., NNVB states come in a
large number of topological sectors characterized by
two integer winding numbers $(n_x,n_y)$. (For a review, see e.g. Ref. \onlinecite{ML}). 
When the same lattice is viewed as a rectangle with open boundary
conditions, the remaining allowed NNVB state all belong to 
a {\em subset} of just the $(0,0)$ sectors. In the thermodynamic limit,
the number of NNVB states for open boundary conditions
thus becomes a vanishing fraction of the corresponding number for periodic
boundary conditions.
It is thus clear that the statement of linear independence
becomes considerably stronger for periodic boundary conditions,
and is often desirable in applications.

We proceed by applying and refining a method that has recently
been developed for the kagome lattice,\cite{a_seidel09} making it
amenable to more general lattice structures.
In Section \ref{derivation} we review this method.
In Section \ref{lattices} we report that this method
can be applied  without much alteration
to the honeycomb lattice, the star lattice,
the square-octagon lattice, the squagome lattice, two types of pentagonal lattices
(studied in a magnetic context, e.g., in Refs. \onlinecite{raman1} and \onlinecite{decorate}), 
three types of ``martini'' lattices,\cite{ziff} and two types of archimedean lattices.
In Section \ref{C60}, we apply the same method to
fulleren-type lattices.
We find that the case of the square lattice  requires a generalization of this method,
which is introduced and applied in Section \ref{squarelattice}.
In Section \ref{summary} we summarize our results
and discuss possible further applications.

\section{Method and results}
\subsection{Derivation of the linear independence condition}\label{derivation}
In this section we review the method used in Ref. \onlinecite{a_seidel09} to prove the linear independence of the 
nearest neighbor valence bond states on the kagome lattice. 
We find that this method can be extended straightforwardly 
to most other lattices to be considered here. A refinement
necessary to study the case of the square lattice will be given
further below.

\begin{figure}[t]
\centering
\includegraphics[width=8cm]{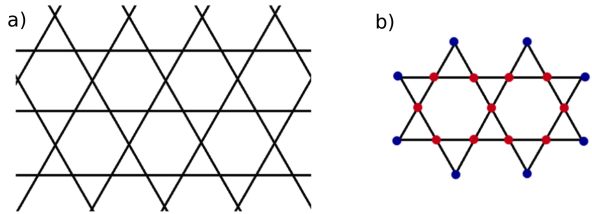}
\caption{
(Color online.) The kagome lattice.
a) shows the structure of the kagome lattice, while b) shows the minimal (smallest) cell for 
which the local independence property defined in the text was proven.\cite{a_seidel09} Different dots 
are used to label the sites which are defined as inner and outer sites, respectively.\label{kagome}
 }
\end{figure}

The general starting point of this method is the identification of a 
suitable (ideally, smallest) cell for which a rather strong local linear independence 
property  holds true.
This local linear independence property can conveniently be verified
numerically, although in many cases an analytic proof seems feasible 
as well. As shown in Ref. \onlinecite{a_seidel09}, this local property
then implies the linear independence of nearest neighbor valence bond
states on arbitrarily large lattices that can, in a certain sense, be covered
by such cells.\cite{note1}
To make this paper self-contained,
we will repeat the proof in the following.
For the kagome lattice,
the smallest possible  cell 
that satisfies these requirements 
is the 
19-site ``double star'' shown in \Fig{kagome}.

For any given cell of a lattice, we define as interior or inner sites of the cell those sites
for which all nearest neighbors are also contained within the cell.
Here, the nearest neighbors of a site are all sites connected to it through a link of the lattice.
Sites that are not interior are called the boundary sites of the cell.
For the kagome cell depicted in \Fig{kagome}, all sites belonging to one of the
internal hexagons are interior, while the remaining ones are boundary sites, unless
the cell happens to be at a boundary of the lattice itself.
In this work we will, however, mostly consider lattices without boundary.
Statements about lattices with boundary can then be obtained as simple corollaries.
Therefore,
the distinction between interior and boundary sites within a cell such as
shown in \Fig{kagome} will not depend on the position of the cell within the lattice.

\begin{figure}[t]
\centering
\includegraphics[width=8cm]{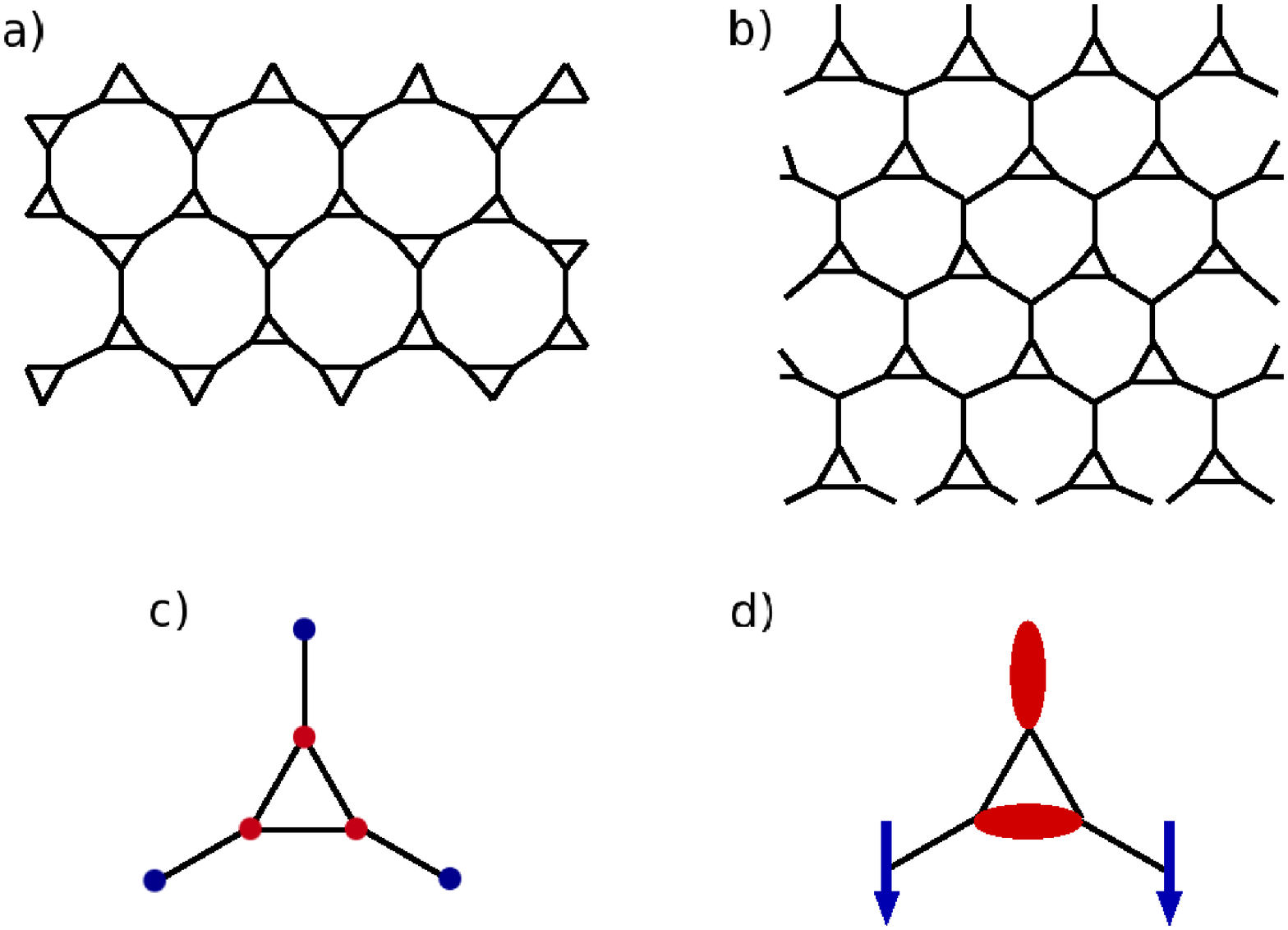}
\caption{(Color online.)  The star lattice (a), and its minimal cell (c) for which the local independence
property could be established. (b) shows the martini-A lattice, with the same minimal cell (c).  Different shades (colors) of dots identify internal and boundary sites.
(d) shows a possible dimer covering: the internal sites must be touched by a dimer, 
boundary sites may or may not form a dimer (valence bond) with an internal site.
In a local valence bond state,
boundary sites not participating in valence bonds may be in an arbitrary spin configuration.
 \label{star_cellstuff}
 }
\end{figure}

To proceed, we will now define a certain class of states living on the local cells. 
We will refer to these states as ``local valence bond states''.
This does, however, not imply that these states completely dimerize the cell, i.e. that every site of the cell 
must participate in a valence bond within the cell. 
Rather, we think of these states as local ``snapshots'' of a lattice
that is in a (globally defined) nearest neighbor valence bond state.
In such a snapshot, every internal site of the cell must certainly
form a valence bond with one of its nearest neighbors
within the cell. A boundary site of the cell, however, may
or may not participate in a valence bond with a site 
within
the cell under consideration. In particular, it may participate
in a valence bond with a site {\em outside} that cell.
In the latter case, the local density matrix describing the
state of the cell contains no information about the state of the
spin of such a boundary site. This motivates the
following definition of local valence bond ``snapshot''
states on the cell $\cal C$. 
Let us consider states of the form
\begin{equation}\label{snapshot}
   |D\rangle \times |\psi_f\rangle\;.
\end{equation}
Here, $D$ represents a dimer covering of the cell $\cal C$.
By this we mean a pairing of the sites of the
cell $\cal C$ into nearest neighbor pairs, where each internal site
is a member of a pair, but not necessarily each boundary site.
An example for such a pairing is given for the cell 
of the star lattice shown in Fig. (\ref{star_cellstuff}d),
and that of the square lattice shown in Fig. (\ref{SQUARE3}c).
By $|D\rangle$ we denote a state where each pair of $D$ forms a singlet, with an arbitrary phase
convention. In \Eq{snapshot}, the state $|\psi_f\rangle$  then denotes any state of the ``free'' sites that
are left untouched by the dimer covering $D$.
This can again be seen in  Figs. (\ref{star_cellstuff}d), (\ref{SQUARE3}c). 
In (\ref{SQUARE3}c), every dimer covering $D$ leaves behind at
least one free site, because of the odd number of sites in this cell.
For cells of even size, we leave it understood that the factor $|\psi_f\rangle$ in 
\Eq{snapshot} is absent if $D$ covers all sites of the cell.

We find it convenient to denote by $\cH(D,\cC)$ the linear space
formed by all local states of the form \eqref{snapshot}, for a {\em fixed}
dimer covering $D$, and will also write $\cH(D)$ instead of
$\cH(D,\cC)$ whenever it is clear what cell is being referred to.
The space spanned by all states of
this form, without fixing $D$, is called the local valence bond
space of the cell $\cC$, $VB(\cC)$:
\begin{equation}\label{VB}
VB(\cC)= \sum_D \cH(D,\cC)\;.
\end{equation}
Here, the sum denotes the linear span. For a given cell
$\cC$, we will now ask whether the sum in \Eq{VB} is {\em direct}.
This means that the expansion of any state in $VB(\cC)$
into members of the various spaces $\cH(D)$ is possible
in one and only one unique way.
Whenever this property holds for some cell $\cC$, we will say
that the NNVB states are ``locally
independent'' on the cell $\cC$, or satisfy the
``local independence property'' on the cell $\cC$.

The local independence property, whenever it can be established
for some cell $\cC$, extends to arbitrarily large lattices that can be
covered by cells of this topology. Said more precisely,
we require that every link of the lattice belongs to a cell 
that has the topology of $\cC$.\cite{note2}
The linear independence of NNVB states defined on the
entire lattice can then be seen as follows.\cite{a_seidel09}
If the sum in \Eq{VB} is direct, then linear projection
operators $P_D$ acting on the cell $\cC$
are well defined, which project onto the subspaces 
$\cH(D)$. Said differently, the defining properties of these
operators are
\begin{equation}\label{PD}
  \begin{split}
     P_D\;\ket{D'}\otimes\ket{\psi_f}&=\delta_{D,D'}\,\ket{D'}\otimes\ket{\psi_f}\,,\\
     \text{hence} \;\; P_D P_{D'}&= \delta_{D,D'} P_D\;.
  \end{split}
\end{equation}
We note that since the spaces $\cH(D)$ are not orthogonal, the linear
projection operators thus defined are not Hermitian.
We also mention that to define these operators in within the full
$2^{|\cC|}$ dimensional Hilbert space of the cell $\cC$,
we need to specify their action on a suitably chosen complement of the
local valence bond space $VB(\cC)$, which can be done in an arbitrary way.
In the following, we will only need to know the action of these operators
within the subspace $VB(\cC)$.

\begin{figure}[b]
\centering
\includegraphics[width=8cm]{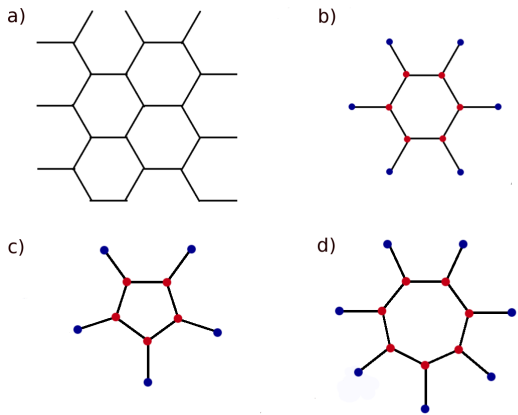}
\caption{
(Color online.) Honeycomb and related structures. (a) The honeycomb lattice. (b) its  minimal cell  with internal and boundary sites identified. (c) The minimal cell of the buckyball lattice (\Fig{bucky}).
(d) A similar heptagonal cell that also satisfied the local independence property.
 \label{honey}
 }
\end{figure}

The operators $P_D$ can now be defined for any cell $\cC$ of some lattice $\cL$,
for which the nearest neighbor valence bond states are locally independent
in the sense defined above. We may write $P^\cC_D$ to explicitly refer
to the cell $\cC$ on which these operators act, but will continue to write 
$P_D$ instead whenever no confusion is possible. Armed with these
operators, we may consider a general linear relation
of the form 
\begin{equation}\label{lc}
\sum_{D'} \lambda_{D'} |D'\rangle =0 \;.
\end{equation}
Here, $D'$ now represents a full dimerization of the entire lattice,
and for simplicity, we assume that the lattice has no boundary,
and can be covered by a single type of cell, as defined above. 
We will comment on the (simpler) case where the lattice
has a boundary below. The states $|D'\rangle$
are thus NNVB states 
of the lattice $\cL$.
For definiteness, we may think of, e.g., a honeycomb lattice
with periodic boundary conditions.
The honeycomb lattice and its smallest cell for which the
local independence property holds are shown in \Fig{honey}.
We want to show that \Eq{lc} implies that all coefficients $\lambda_{D'}$
are zero.
For this we first focus on a single cell $\cC$ of the lattice that has the topology
shown in Fig. (\ref{honey}b), and a fixed dimer covering $D$ of
the entire lattice. The dimer covering $D$ determines a dimer covering $D_\cC$
of the cell $\cC$, consisting of those dimers of $D$ that are fully contained in $\cC$.
Consider the action of the operator $P_{D_\cC}$ defined for the cell $\cC$
on any of the states $|D'\rangle$ in \Eq{lc}.
Clearly, the dimer covering $D'$ determines a local dimer covering of $\cC$,
$D'_\cC$, defined analogous to $D_\cC$. From the definition of the projection operators,
\Eq{PD}, we see that
\begin{equation}\label{PDglobal}
P_{D_\cC}|D'\rangle = \delta_{D_\cC,D'_\cC} |D'\rangle\,.
\end{equation}
This is so since the state $|D'\rangle$ is 
contained
in the tensor product $\cH(D'_{\cC},\cC)\otimes\cH(\cL\setminus\cC)$,
where the second factor denotes the Hilbert space 
associated with all lattice sites not contained in $\cC$.
$P_{D_\cC}$ only acts on the first factor, and does so according to
\Eq{PD}.
Some further (but trivial) details are explicitly written in Ref. \onlinecite{a_seidel09}.
Hence, when $P_{D_\cC}$ acts on \Eq{lc},
one obtains a similar linear combination on the left hand side,
but with all dimer coverings $D'$ omitted for which
the cell $\cC$ does not contain {\em exactly the same} dimers
as for $D$.
We can proceed by successively acting on this new linear relation
with the operators $P_{D_{\cC'}}$, where $D$ is the same as before,
but $\cC'$ now runs over all cells of the lattice with the same
topology as $\cC$. Since by assumption, these cells cover the lattice
in the sense defined above, only those states $|D'\rangle$
in \Eq{lc} survive this procedure whose underlying dimer covering
$D'$ looks the same as $D$ everywhere, i.e., only the term with $D'=D$
survives.
The resulting equation is thus $\lambda_D |D\rangle=0$,
which implies $\lambda_D=0$. 
Hence $\lambda_{D'}=0$ for each dimer covering $D'$,
since $D$ was arbitrary. This then proves the linear independence
of the nearest neighbor valence bond states on the lattice $\cL$.

So far we have considered lattices with periodic boundary conditions.
The above result, however, immediately carries over to
lattices with a boundary. Let us consider any lattice $\cL'$ with an edge that can be obtained from a lattice
$\cL$ with periodic boundary conditions,
for which the linear independence of NNVB states has been proven,
 by means of the removal of 
certain boundary links.
Then the set of full dimerizations $D$ of $\cL'$ is just a subset of those of $\cL$,
and likewise the corresponding  set of NNVB states. Hence, if the
linear independence of NNVB states
holds for $\cL$, it must also hold for $\cL'$.
More generally, it is easy to see that our result applies to any
 sublattice $\cL'$ of $\cL$, such that $\cL=\cL'\cup\cL''$
 is a disjoint union, and both $\cL'$ and $\cL''$ are fully dimerizable.

\subsection{Twelve different 2D lattices} \label{lattices}

We now discuss the applicability of this method to various two-dimensional lattices.
As discussed in Section \ref{derivation}, this merely requires the identification of
a cell of the lattice, for which the local independence property holds, and which can
cover
the entire lattice in the sense defined there.
Such cells 
have also been dubbed ``bricks of linear independence'' in Ref. \onlinecite{a_seidel09}.
For brevity, we will refer to the cells identified by us as ``minimal cells'',
since there are presumably (in some case obviously) no smaller
cells with this property on the respective lattices.
We have, however, not carefully ruled out the existence of smaller 
cells in all cases, since this is of limited interest once sufficiently small
``bricks of linear independence'' have been identified.
For the cell $\cC$ in question, we pick an
appropriate basis $|D\rangle\otimes|\psi_{i}\rangle$ for each space $\cH(D)$, 
where $i=1\dots 2^{n_D}$, and $n_D$ is the number of sites of the cell $\cC$ 
that do not participate in
the local dimer recovering $D$.
The local independence property introduced in the preceding section is then equivalent to the statement
that the overlap matrix 
\begin{equation}\label{matrixelement}
M_{D',j;D,i}=(\langle D'|\otimes\langle \psi_j|) \,(|D\rangle\otimes|\psi_i\rangle)
\end{equation}
has full rank. It is clear that for a suitable choice of the factors $|\psi_i\rangle$, e.g. 
``Ising''-type basis states with well-defined local $S_z$, and suitable overall normalization
factors, the matrix elements
$M_{D',j;D,i}$  are integer. The question of the rank of this matrix can 
then be addressed using integer arithmetic free of numerical errors.
We did this by using the LinBox package.\cite{LinBox}
By choosing the $\psi_i$ from an Ising-$S_z$ basis, the matrix in \Eq{matrixelement}
is also block diagonal with blocks of definite total $S_z$.
This let to manageable matrix sizes in all the cases discussed in this section.

\begin{figure}[t]
\centering
\includegraphics[width=8cm]{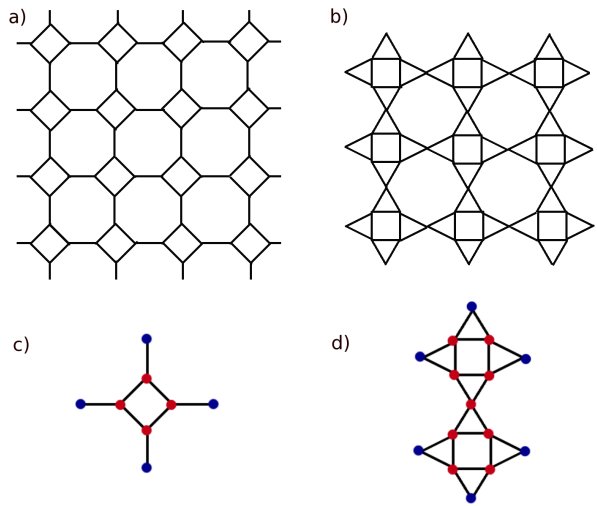}
\caption{
(Color online.) The square-octagon lattice (a) and the squagome lattice (b). 
(c) and  (d) show the respective minimal cells.\label{square_octagonal_squagome}
 }
\end{figure}

\begin{figure}[t]
\centering
\includegraphics[width=8cm]{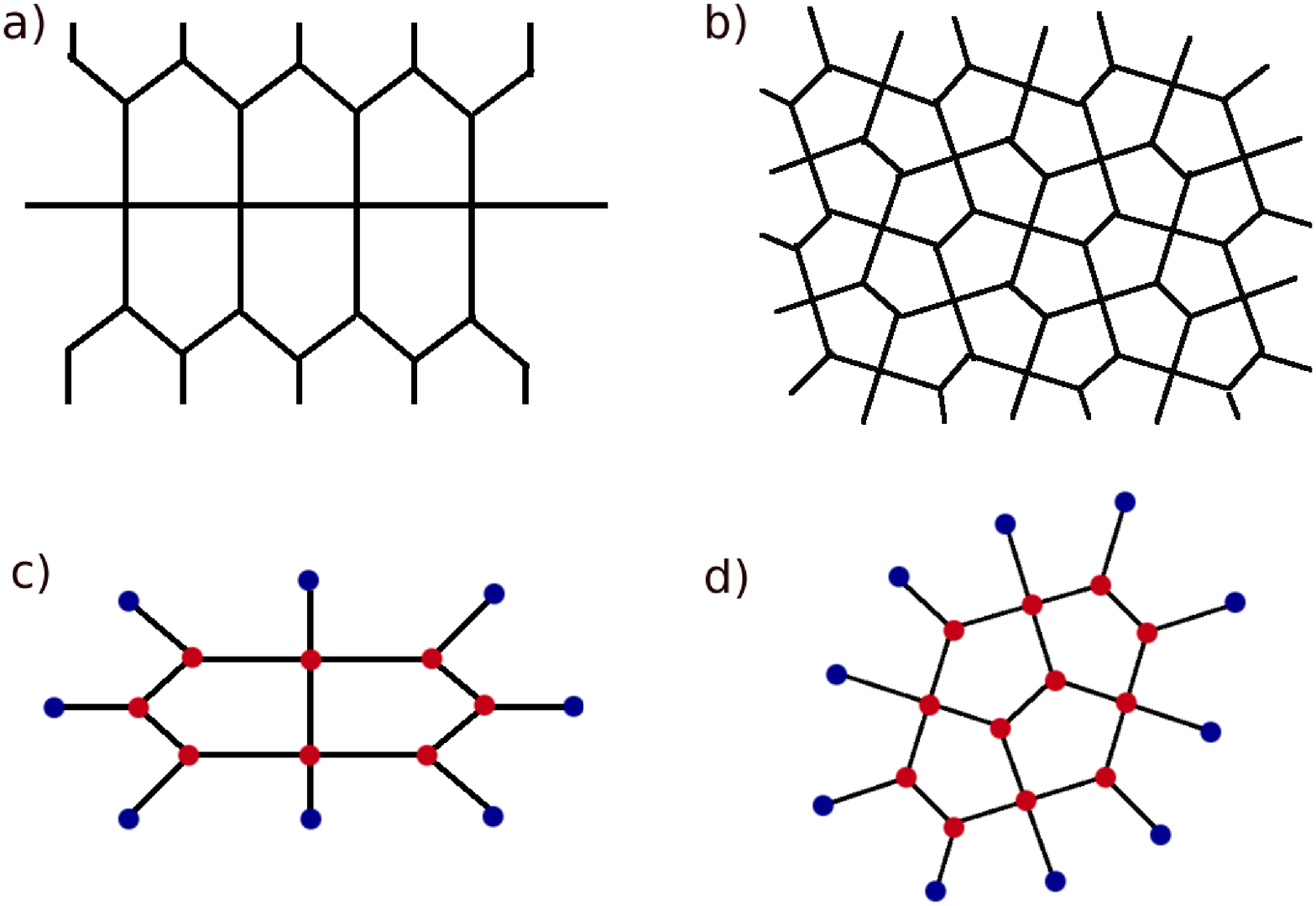}
\caption{
(Color online.) Two types of pentagonal lattices.
(a) shows the pentagonal lattice and (b) shows the ``Cairo'' pentagonal lattice structure,
(c) and (d) the respective minimal cells. \label{penta_I_II}
 }
\end{figure}

\begin{figure}[t]
\centering
\includegraphics[width=8cm]{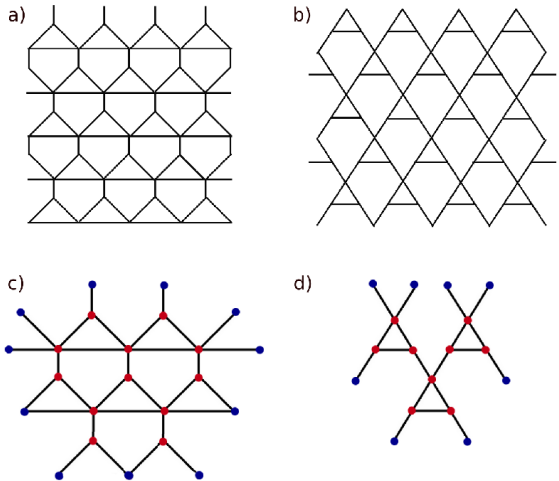}
\caption{
(Color online.) Two more types of martini lattices.
(a) and (b) show the lattice structures of the martini-B lattice and the martini-C lattice, respectively, 
(c) and (d) the respective minimal cells. \label{martini_middleright}
 }
\end{figure}

\begin{figure}[t]
\centering
\includegraphics[width=8cm]{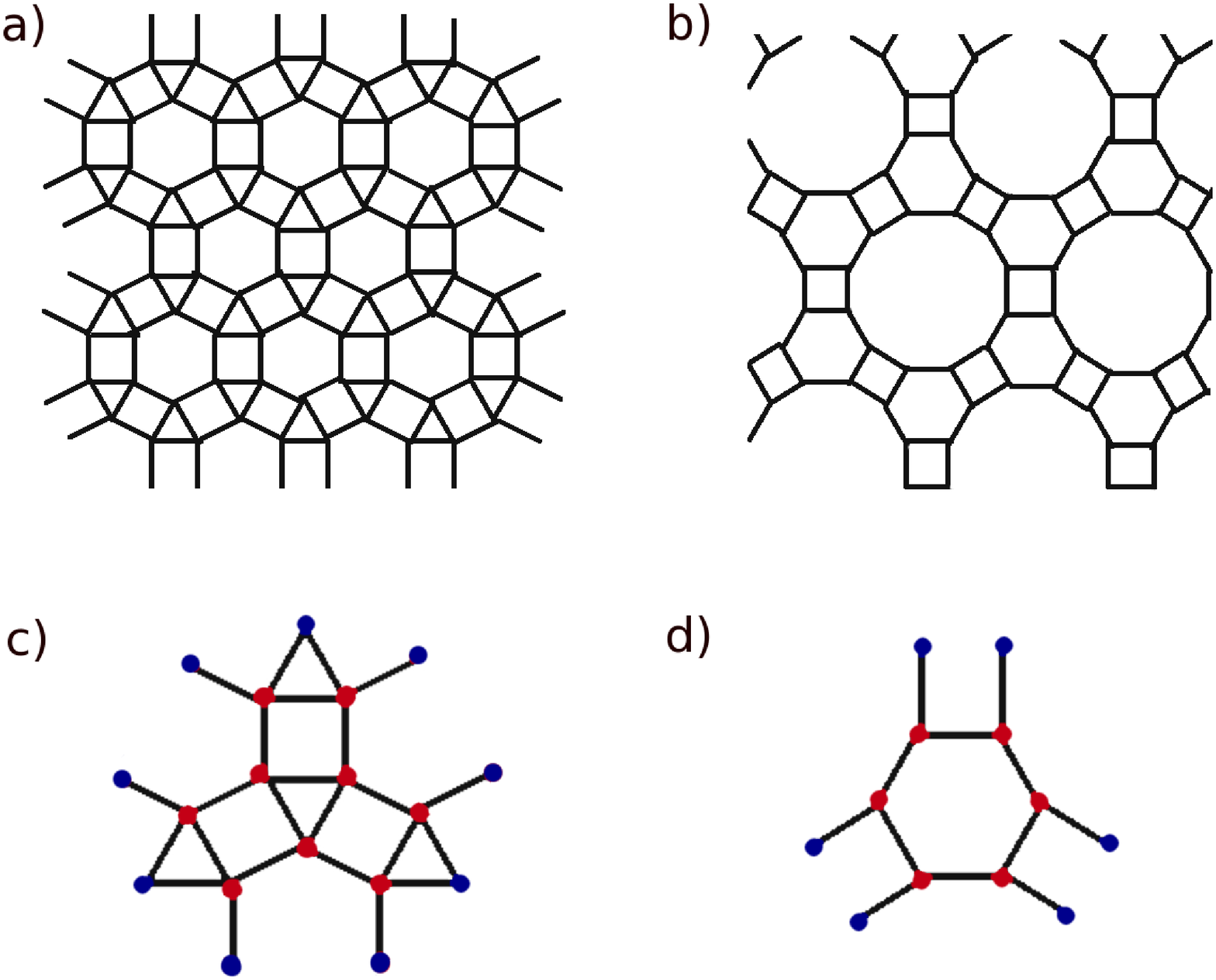}
\caption{
(Color online.) Two types of archimedean lattices.
(a) shows the archimedean-A lattice and (b) shows the archimedean-B lattice structures,
(c) and (d) the respective minimal cells. Note that the minimal cell in (d) is the same as 
that of the honeycomb lattice, Fig. (\ref{honey}b).\label{arch}
 }
\end{figure}

We present twelve different 2D lattices which we successfully 
studied using the method described above, and their respective minimal cells $\cC$,
 for which the local independence has been found to hold, Figs. \ref{kagome}-\ref{arch}. 
These are, in order, the kagome lattice (treated in Ref. \onlinecite{a_seidel09}),
the star lattice, the martini-A lattice, the honeycomb lattice, the square-octagon lattice, the squagome lattice, 
the pentagonal and the ``Cairo'' pentagonal lattice, 
2 more types of the``martini'' lattice (martini-B and martini-C), 
and two types of so called archimedean lattices, denoted archimedean-A and archimedian-B.  
As proven above, for all these lattices, the identification 
proper ``bricks of linear independence''
implies the linear independence NNVB states for arbitrarily large lattices of this type (which must also
be
large enough to contain the minimal cell), 
for both open and periodic boundary conditions.
For the square-octagon and the honeycomb lattices, 
the case of open boundary 
conditions had already 
been treated
in Ref. \onlinecite{kcc} by a different method. 
It is interesting to note that the size of the matrix in \Eq{matrixelement} differs
quite significantly for the 2D lattices discussed here: for the star and the martini-A lattice,
which share the same minimal cell
(\Fig{star_cellstuff}), the total matrix dimension
(over all $S_z$-blocks)  is only 13.  For others, the matrix dimension
is on the order of a few thousand, and for the square lattice cell treated separately in Section \ref{squarelattice}, the set of ``local valence bond'' states
$|D\rangle\otimes |\psi_i\rangle$ defining the matrix has more than half a million elements.


\begin{figure}[t]
\centering
\includegraphics[width=6cm]{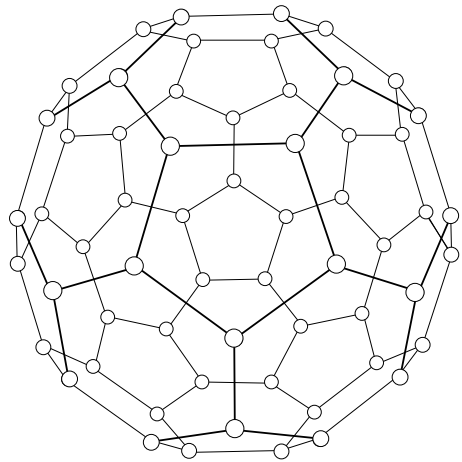}
\caption{\label{bucky}
The lattice of the C$_{60}$ molecule, or ``buckyball''.
The lattice can be covered by the minimal cell of the
honeycomb lattice,  Fig. (\ref{honey}b).
The actual minimal cell of this lattice is the pentagonal cell shown
in Fig. (\ref{honey}c).
 }
\end{figure} 

\subsection{Fullerene-type lattices}\label{C60}

We now consider fullerene-type lattices,
where each site has three nearest neighbors, and belongs to at least one
hexagonal plaquette, where no two members of the same hexagonal
plaquette share a nearest neighbor outside that plaquette.
Such lattices can be covered, in the sense defined in
Section \ref{derivation}, by the minimal cell of the honeycomb lattice,  Fig. (\ref{honey}b).
A famous example is the ``Buckyball'' lattice, \Fig{bucky}.
By the results of the preceding sections, the NNVB states 
on these types of lattices are
are linearly
independent.  
This also demonstrates that our method, being essentially local,
  can be applied to general
 lattice topologies. \cite{note3} 

 The Heisenberg model on fulleren-type lattices has
 been  extensively studied within the NNVB  
 subspace  in Ref. \onlinecite{klein} (there called Kekul\'e subspace). Good agreement
 with exact diagonalization results within the full Hilbert space
 was found. The authors also point out the central importance of the linear
 independence of the NNVB states to their approach. 
 Since the fullerene lattices are finite in size,
 conventional brute-force numerics may in principle be used to establish
 this, although the feasibility of this depends, of course,
 on the actual lattice size. In contrast, the result derived here holds for
 arbitrarily large systems, and, given the small size of the minimal
 cells involved, could be obtained purely analytically.
In this regard, it is worth noting that the actual {\em minimal} cell of the
 C$_{60}$ molecule is not that of the honeycomb lattice, Fig. (\ref{honey}b),
 but the smaller pentagonal cell of Fig. (\ref{honey}c).
 We have verified that it likewise satisfies the local independence
 property, and each link of the Buckyball lattice belongs to such a cell.
 For this rather small cell, an analytic proof of the local
 independence property seems feasible
   using the Rumer-Pauling method \cite{rumer,pauling,soos1,soos2}
 referred to in the next section.

 Based on the observations made thus far,
 we conjecture that all cells where the $n$ inner sites form
 a single polygon, and each inner site is 
 linked to exactly one of $n$ boundary sites, have the local independence
 property. Examples of such cells, for which we have verified this,
 are given in Figs. (\ref{star_cellstuff}c), (\ref{square_octagonal_squagome}c),  and (\ref{honey}b-d), 
 corresponding to $n=3,4,5,6,7$.
For all lattices that can be covered by any combination of such cells (see Ref. \onlinecite{note1}),
we thus have the linear independence of NNVB states.

\subsection{The square lattice}\label{squarelattice}

\begin{figure}[t]
\centering
\includegraphics[width=8cm]{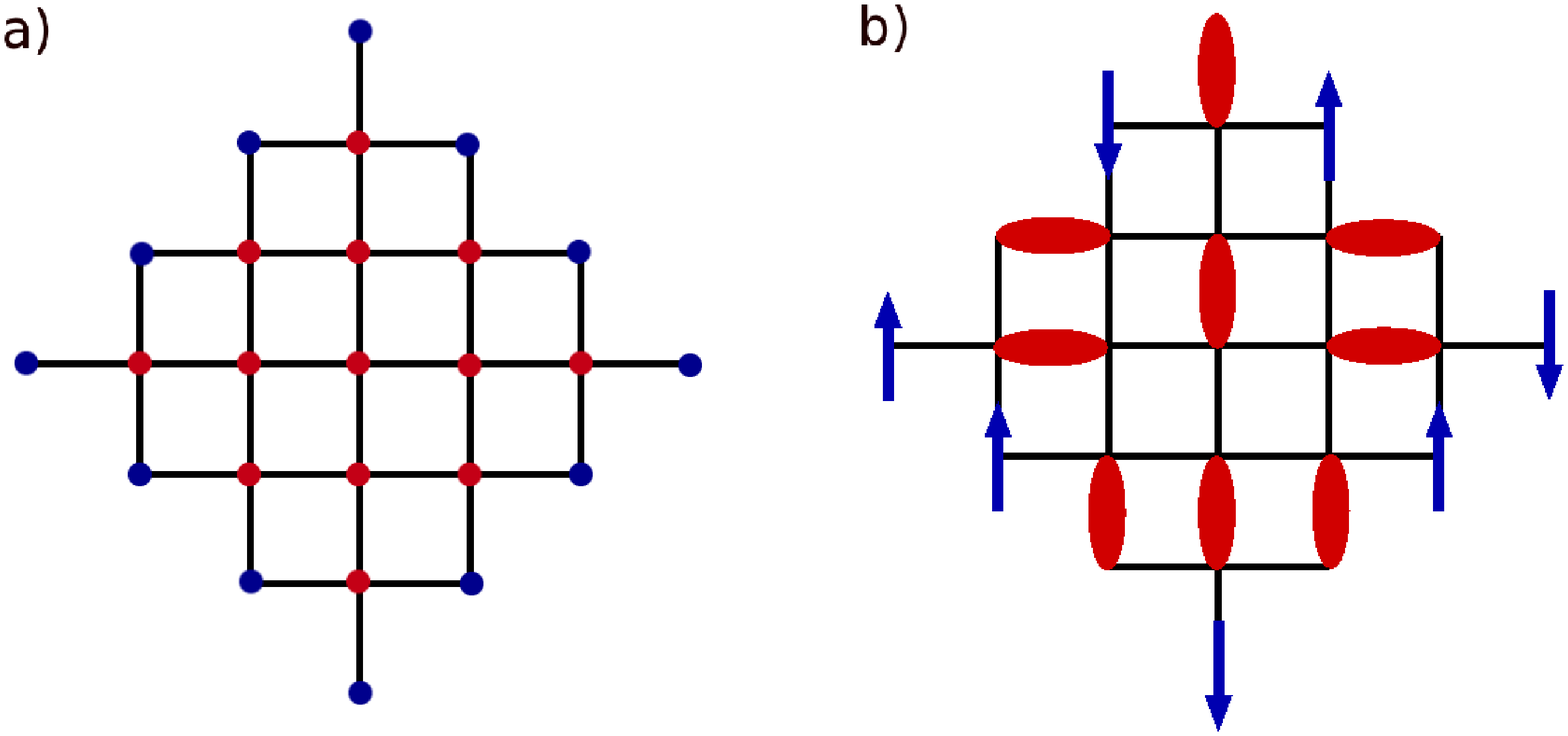}
\caption{\label{SQUARE3} (Color online.) 
The square lattice.
The general lattice structure is shown in \Fig{dimers}.
(a) The minimal cell for which the
refined local independence property
of Section \ref{squarelattice} holds.
(b) A local valence bond state with the central site forming a
bond with its upper neighbor, corresponding to $\sigma=\uparrow$ as defined in the text.
 }
\end{figure}
We find that the method presented in Sec. \ref{derivation} cannot immediately be applied to the square lattice.
The problem can be traced back to the fact that any local cell on this lattice necessarily that 90 degree
corners. It turns out that by using the degrees of freedom near these corners, one can always form non-trivial
relations between the states in different subspaces $\cH(D)$. 
The projection operators in \Eq{PD} are then ill-defined.
We thus have to modify our method in order
to deal with this case. Luckily, the local independence property introduced in Sec. \ref{derivation},
while it is found to hold for many lattice types, is overly restrictive.
In fact, whenever this property holds,
it can be literally extended to arbitrarily large lattices with an edge.\cite{a_seidel09}
That is, for an arbitrarily large lattice $\cL$,
 not only states $|D\rangle$ corresponding to full dimerizations of $\cL$ are then 
linearly independent, but in fact all states of the form $|D\rangle\otimes|\psi_i\rangle$,
where $D$ does not necessarily cover all boundary sites of $\cL$, and the factors $|\psi_i\rangle$
form a basis of the space associated with ``free'' boundary sites. 
Clearly, this is a stronger statement than just the linear independence of NNVB states
corresponding to full dimerizations of the lattice.
However, 
for the square lattice this stronger property simply does not hold. 
On the other hand, this ``strong'' linear independence property is not   
of primary interest. We are still interested in the linear independence of NNVB states associated with full
dimerizations  of the lattice, for which the stronger property is not necessary.
 
It turns out that a weaker version
of the local independence property is sufficient to construct suitable projection operators
for our purpose. To see this, note that
the operators $P_D$ defined in \Eq{PD} are sensitive to the entire configuration of valence
bonds fully contained within the cell on which they act. To prove the linear independence
of NNVB states, it is sufficient to work with operators that
are sensitive, say, to the bonding state of any given site, as determined by which nearest neighbor this central
site is bonding with. 
To accomplish this, we consider a square lattice $\cL$ satisfying periodic
boundary conditions, which is large enough to contain the cell $\cC$ depicted in  Fig. (\ref{SQUARE3}b).
For this cell, we consider four subspaces of $VB(\cC)$, according to the bonding state
of the central site.  We define local dimer coverings $D$ of $\cC$ as before, where
boundary sites of $\cC$ need not be covered. By $\sigma(D)$ we denote the bonding state
of the central site, i.e., $\sigma(D)=\uparrow,\downarrow,\leftarrow,\rightarrow$,
depending on whether this site is paired with its upper, lower, left, or right neighbor in the
covering D, respectively. 
As mentioned initially, the sum \Eq{VB} defining the space of local 
valence bond states, $VB(\cC)$, is not direct for the present cell. However, we may 
also write the space $VB(\cC)$ as a ``courser'' sum of fewer spaces,
each of which is formed by all valence bond states that have the
central site in a certain bonding state:
\begin{equation}\label{VBcourse}
   VB(\cC)= \sum_{\sigma'}   VB_{\sigma'}(\cC)\;,
\end{equation}
where
\begin{equation}
VB_{\sigma'}(\cC)=\sum_{\substack{D\\ \sigma(D)=\sigma'}} \cH(D,\cC)\;,
\end{equation}
and $\sigma'$ runs over all possible values $\uparrow,\downarrow,\leftarrow,\rightarrow$.

The key observation that renders the square lattice amenable to our method is that the
sum in \Eq{VBcourse} is still direct. To show this, one must show that the dimensions
of the spaces on the right hand side add up to the dimension of the space
$VB(\cC)$. For this it is sufficient to consider the matrix 
$M$ defined in 
\Eq{matrixelement}, together with the matrices $M_{\sigma'}$ that are the submatrices
of $M$ corresponding to the subspaces $VB_{\sigma'}(\cC)$, and show that the
ranks of the $M_{\sigma'}$'s add up to that of $M$.
Intuitively speaking, this means that while the states $|D\rangle\otimes |\psi_i\rangle$,
as defined below \Eq{matrixelement}, satisfy non-trivial linear relations, all these
linear relations can be restricted to involve members of the same subspace
$VB_{\sigma'}(\cC)$; there are then no further linear relation between members of 
different subspaces. 
If the sum in \Eq{VBcourse} is indeed direct, we may introduce projection operators
$P_{\sigma'}$ onto the components on the right hand side. The defining
property of these operators is
\begin{equation}\label{Psigma}
     P_{\sigma'}\;\ket{D}\otimes\ket{\psi_i}=\delta_{\sigma',\sigma(D)}\,\ket{D}\otimes\ket{\psi_i}\,.
\end{equation}
When acting on local valence bond states $\ket{D}\otimes\ket{\psi_i}$ living on the cell $\cC$, the operator
$P_{\sigma}$ will thus annihilate the state if the bonding state of the central site in the dimer
covering $D$ is different from $\sigma$, and otherwise leave the state invariant.
It is clear that any site $i$ of the periodic (and sufficiently large) lattice $\cL$ can be made the central site 
of a cell that has the topology of $\cC$, Fig. (\ref{SQUARE3}b). The operators $P_{\sigma}$ defined above can then be 
extended to the full Hilbert space of the large lattice, and there is an operator
$P^i_{\sigma}$ for any cell of the type $\cC$ with central site $i$.
The defining property \eqref{Psigma} then extends to valence bond states $|D\rangle$
corresponding to full dimerizations $D$ of the lattice: $|D\rangle$ survives the
action of $P^i_{\sigma}$ unchanged if the bonding state of site $i$ in the
dimer covering $D$ is $\sigma$, otherwise it is annihilated. Detailed arguments
for this are identical to those referred to in Sec. \ref{derivation}.
It is then clear that by successive action with the operators $P^i_\sigma$,
one can single out any dimer covering $D$ in the linear combination \Eq{lc},
just as carried out in Section \ref{derivation},
and thus prove that the states $|D\rangle$ are linearly independent.

We have verified that for the cell in Fig. (\ref{SQUARE3}b), the sum in \Eq{VBcourse}
is indeed direct. The numerics were somewhat more challenging, due to size
of the 25-site cell under consideration. To wit, this cell admits a 
total of 5376 different dimer coverings.
Each of the dimer coverings has seven ``free'' outer sites not touched by a dimer, thus the total 
dimension of the $M$-matrix
 is a staggering 5376 x 2$^7$ = 688128. To reduce the problem to blocks
 of manageable size, we use the full rotational invariance of the
 spaces appearing in \Eq{VBcourse}. That is, we chose the basis $|\psi_i\rangle$
 for the seven free sites to have a well-defined total spin $S$, in addition
 to a well-defined $S_z$. A suitable choice for a basis is obtained
  by choosing states corresponding to Rumer-Pauling diagrams.\cite{rumer,pauling,soos1,soos2}
  The advantage of this is that for appropriate normalization, the matrix
  elements of the $M$-matrix then remain integer, and we may again make
  use of exact integer arithmetic.\cite{LinBox} We further used the mirror
  symmetry of the cell $\cC$ along one of its diagonals. The largest blocks
  occurring then had dimensions on the order of 30,000.
  
 The above then establishes that for {\em any} sufficiently large
 square lattice with periodic boundary conditions, the set of all
NNVB states is linearly independent.
 The same statement then follows for lattices with an edge
 as discussed at the end of Section \ref{derivation}. 
 The case of general open boundaries conditions has also been treated previously
 with different methods
 in Ref. \onlinecite{kcc}.  
  
\section{Summary and discussion}\label{summary}
In the preceding sections, we have described a method for proving linear independence of nearest neighbor
valence bond states on certain 2D lattices with and without periodic boundary conditions. 
This method, originally designed for the kagome lattice,\cite{a_seidel09} was successfully extended here to the following lattice types: 
honeycomb, squagome, pentagonal and Cairo pentagonal, square-octagon, martini-A, -B, and -C, 
 archimedian-A and archimedean-B, and to the star lattice,
 and furthermore to fullerene-type lattices.
Subsequently, a refined method has been developed, which is applicable even in some cases where the original
method is inadequate. Specifically, this was found to be the case for the square lattice.

Our method is based on the identification of a certain local independence property for finite clusters,
which, when established, implies the linear independence of NNVB states
for arbitrarily large lattices. Though here we prefer to validate the local independence property
using exact numerical schemes, in those cases where smaller clusters are sufficient,
a fully analytic approach is certainly feasible. Further remarks on this for the kagome case,
where the cluster size is fairly large, can be found in
 Ref. \onlinecite{a_seidel09}.
 
We note again that the linear independence of the NNVB states for the square, the honeycomb, and the square-octagon 
lattice was already established in a paper by Chayes, Chayes, and Kivelson\cite{kcc} in 1989.
Their result, however, applies only to the case of open boundary conditions. 
For these lattices, our result extends the one by Chayes et al. to the case of periodic boundary conditions, using a
different approach. We have also discussed various applications of these results
in RVB inspired approaches to  quantum spin-$1/2$ systems.

A case of much interest, which we have not studied here, is that of the triangular lattice.
We remark that since the square lattice can be 
thought of as a triangular lattice endowed with a coarser topology,
obtained by removing certain nearest neighbor links, a candidate
cell for the triangular lattice would have to be at least as large as our square
lattice cell, Fig. (\ref{SQUARE3}b), with many more links included.
This renders the $M$-matrix so large that we did not find this 
problem tractable at present. We currently see, however,
no fundamental reason why the refined method
of the preceding section should not be applicable to this 
case as well. In all cases thus far studied,  we have found that
local cells large enough to have {\em more}
internal sites than boundary sites generally have
a sufficiently strong local independence property,
which then implies the desired linear independence of globally
defined NNVB states.
The only exception to this rule seem to be lattices where this ``global''
linear independence does not hold,  for obvious, ``local'' reasons:
These include the checkerboard and the pyrochlore lattice,
or any lattice featuring tetrahedral units. By looking at the three dimerizations of a single
tetrahedron, it is easy to see that for such lattices,
linear independence of NNVB states
does not hold. (That is, as long as there is any
dimer covering of the lattice with two dimers on the same tetrahedron.)

Thus far, we are not aware of rigorous results on the problem
studied here for any three-dimensional lattices (except for finite clusters).
We are optimistic, however, that our method is at least applicable
to the  hyperkagome case, which has recently enjoyed much attention
in the study of frustrated quantum magnets.\cite{hyperkg_experiment,hyper1,hyper2,hyper3,hyper4,hyper5}
A brute force study of the relevant local cell has so far been barred
by its size. However, a formal analogy with the kagome case suggests
that a partially analytic treatment of the local cell is possible.\cite{a_seidel09}
We leave this and other unexplored cases of interest for future studies.

\begin{acknowledgements}
We would like to thank Z. Nussinov for insightful discussions.
This work has been supported
by the National Science Foundation under NSF Grant No. DMR-0907793.
\end{acknowledgements}





\begin{thebibliography}{86}
\expandafter\ifx\csname natexlab\endcsname\relax\def\natexlab#1{#1}\fi
\expandafter\ifx\csname bibnamefont\endcsname\relax
  \def\bibnamefont#1{#1}\fi
\expandafter\ifx\csname bibfnamefont\endcsname\relax
  \def\bibfnamefont#1{#1}\fi
\expandafter\ifx\csname citenamefont\endcsname\relax
  \def\citenamefont#1{#1}\fi
\expandafter\ifx\csname url\endcsname\relax
  \def\url#1{\texttt{#1}}\fi
\expandafter\ifx\csname urlprefix\endcsname\relax\def\urlprefix{URL }\fi
\providecommand{\bibinfo}[2]{#2}
\providecommand{\eprint}[2][]{\url{#2}}

\bibitem[{\citenamefont{Shastry and Sutherland}(1981)}]{shastrysutherland}
\bibinfo{author}{\bibfnamefont{B.~S.} \bibnamefont{Shastry}} \bibnamefont{and}
  \bibinfo{author}{\bibfnamefont{B.}~\bibnamefont{Sutherland}},
  \bibinfo{journal}{Physica B} \textbf{\bibinfo{volume}{108}},
  \bibinfo{pages}{1069 } (\bibinfo{year}{1981}).

\bibitem[{\citenamefont{Anderson}(1973)}]{anderson1}
\bibinfo{author}{\bibfnamefont{P.~W.} \bibnamefont{Anderson}},
  \bibinfo{journal}{Mater. Res. Bull.} \textbf{\bibinfo{volume}{8}},
  \bibinfo{pages}{153} (\bibinfo{year}{1973}).

\bibitem[{\citenamefont{Anderson}(1987)}]{anderson2}
\bibinfo{author}{\bibfnamefont{P.~W.} \bibnamefont{Anderson}},
  \bibinfo{journal}{Science} \textbf{\bibinfo{volume}{235}},
  \bibinfo{pages}{1196} (\bibinfo{year}{1987}).

\bibitem[{\citenamefont{Shimizu et~al.}(2003)\citenamefont{Shimizu, Miyagawa,
  Kanoda, Maesato, and Saito}}]{triangular1}
\bibinfo{author}{\bibfnamefont{Y.}~\bibnamefont{Shimizu}},
  \bibinfo{author}{\bibfnamefont{K.}~\bibnamefont{Miyagawa}},
  \bibinfo{author}{\bibfnamefont{K.}~\bibnamefont{Kanoda}},
  \bibinfo{author}{\bibfnamefont{M.}~\bibnamefont{Maesato}}, \bibnamefont{and}
  \bibinfo{author}{\bibfnamefont{G.}~\bibnamefont{Saito}},
  \bibinfo{journal}{Phys. Rev. Lett.} \textbf{\bibinfo{volume}{91}},
  \bibinfo{pages}{107001} (\bibinfo{year}{2003}).

\bibitem[{\citenamefont{Itou et~al.}(2008)\citenamefont{Itou, Oyamada, Maegawa,
  Tamura, and Kato}}]{triangular2}
\bibinfo{author}{\bibfnamefont{T.}~\bibnamefont{Itou}},
  \bibinfo{author}{\bibfnamefont{A.}~\bibnamefont{Oyamada}},
  \bibinfo{author}{\bibfnamefont{S.}~\bibnamefont{Maegawa}},
  \bibinfo{author}{\bibfnamefont{M.}~\bibnamefont{Tamura}}, \bibnamefont{and}
  \bibinfo{author}{\bibfnamefont{R.}~\bibnamefont{Kato}},
  \bibinfo{journal}{Phys. Rev. B} \textbf{\bibinfo{volume}{77}},
  \bibinfo{eid}{104413} (pages~\bibinfo{numpages}{5}) (\bibinfo{year}{2008}).

\bibitem[{\citenamefont{Hiroi et~al.}(2001)\citenamefont{Hiroi, Hanawa,
  Kobayashi, Nohara, Takagi, Kato, and Takigawa}}]{kagome1}
\bibinfo{author}{\bibfnamefont{Z.}~\bibnamefont{Hiroi}},
  \bibinfo{author}{\bibfnamefont{M.}~\bibnamefont{Hanawa}},
  \bibinfo{author}{\bibfnamefont{N.}~\bibnamefont{Kobayashi}},
  \bibinfo{author}{\bibfnamefont{M.}~\bibnamefont{Nohara}},
  \bibinfo{author}{\bibfnamefont{H.}~\bibnamefont{Takagi}},
  \bibinfo{author}{\bibfnamefont{Y.}~\bibnamefont{Kato}}, \bibnamefont{and}
  \bibinfo{author}{\bibfnamefont{M.}~\bibnamefont{Takigawa}},
  \bibinfo{journal}{J. Phys. Soc. Jpn.} \textbf{\bibinfo{volume}{70}},
  \bibinfo{pages}{3377} (\bibinfo{year}{2001}).

\bibitem[{\citenamefont{Shores et~al.}(2005)\citenamefont{Shores, Nytko,
  Bartlett, and Nocera}}]{kagome2}
\bibinfo{author}{\bibfnamefont{M.~P.} \bibnamefont{Shores}},
  \bibinfo{author}{\bibfnamefont{E.~A.} \bibnamefont{Nytko}},
  \bibinfo{author}{\bibfnamefont{B.~M.} \bibnamefont{Bartlett}},
  \bibnamefont{and} \bibinfo{author}{\bibfnamefont{D.~G.}
  \bibnamefont{Nocera}}, \bibinfo{journal}{Journal of the American Chemical
  Society} \textbf{\bibinfo{volume}{127}}, \bibinfo{pages}{13462}
  (\bibinfo{year}{2005}).

\bibitem[{\citenamefont{Helton et~al.}(2007)\citenamefont{Helton, Matan,
  Shores, Nytko, Bartlett, Yoshida, Takano, Suslov, Qiu, Chung
  et~al.}}]{kagome3}
\bibinfo{author}{\bibfnamefont{J.~S.} \bibnamefont{Helton}},
  \bibinfo{author}{\bibfnamefont{K.}~\bibnamefont{Matan}},
  \bibinfo{author}{\bibfnamefont{M.~P.} \bibnamefont{Shores}},
  \bibinfo{author}{\bibfnamefont{E.~A.} \bibnamefont{Nytko}},
  \bibinfo{author}{\bibfnamefont{B.~M.} \bibnamefont{Bartlett}},
  \bibinfo{author}{\bibfnamefont{Y.}~\bibnamefont{Yoshida}},
  \bibinfo{author}{\bibfnamefont{Y.}~\bibnamefont{Takano}},
  \bibinfo{author}{\bibfnamefont{A.}~\bibnamefont{Suslov}},
  \bibinfo{author}{\bibfnamefont{Y.}~\bibnamefont{Qiu}},
  \bibinfo{author}{\bibfnamefont{J.-H.} \bibnamefont{Chung}},
  \bibnamefont{et~al.}, \bibinfo{journal}{Phys. Rev. Lett.}
  \textbf{\bibinfo{volume}{98}}, \bibinfo{eid}{107204}
  (pages~\bibinfo{numpages}{4}) (\bibinfo{year}{2007}).

\bibitem[{\citenamefont{Mendels et~al.}(2007)\citenamefont{Mendels, Bert,
  de~Vries, Olariu, Harrison, Duc, Trombe, Lord, Amato, and Baines}}]{kagome4}
\bibinfo{author}{\bibfnamefont{P.}~\bibnamefont{Mendels}},
  \bibinfo{author}{\bibfnamefont{F.}~\bibnamefont{Bert}},
  \bibinfo{author}{\bibfnamefont{M.~A.} \bibnamefont{de~Vries}},
  \bibinfo{author}{\bibfnamefont{A.}~\bibnamefont{Olariu}},
  \bibinfo{author}{\bibfnamefont{A.}~\bibnamefont{Harrison}},
  \bibinfo{author}{\bibfnamefont{F.}~\bibnamefont{Duc}},
  \bibinfo{author}{\bibfnamefont{J.~C.} \bibnamefont{Trombe}},
  \bibinfo{author}{\bibfnamefont{J.~S.} \bibnamefont{Lord}},
  \bibinfo{author}{\bibfnamefont{A.}~\bibnamefont{Amato}}, \bibnamefont{and}
  \bibinfo{author}{\bibfnamefont{C.}~\bibnamefont{Baines}},
  \bibinfo{journal}{Phys. Rev. Lett.} \textbf{\bibinfo{volume}{98}},
  \bibinfo{eid}{077204} (pages~\bibinfo{numpages}{4}) (\bibinfo{year}{2007}).

\bibitem[{\citenamefont{Rokhsar and Kivelson}(1988)}]{RK}
\bibinfo{author}{\bibfnamefont{D.~S.} \bibnamefont{Rokhsar}} \bibnamefont{and}
  \bibinfo{author}{\bibfnamefont{S.~A.} \bibnamefont{Kivelson}},
  \bibinfo{journal}{Phys. Rev. Lett.} \textbf{\bibinfo{volume}{61}},
  \bibinfo{pages}{2376} (\bibinfo{year}{1988}).

\bibitem[{\citenamefont{Moessner and Sondhi}(2001)}]{moessner}
\bibinfo{author}{\bibfnamefont{R.}~\bibnamefont{Moessner}} \bibnamefont{and}
  \bibinfo{author}{\bibfnamefont{S.~L.} \bibnamefont{Sondhi}},
  \bibinfo{journal}{Phys. Rev. Lett.} \textbf{\bibinfo{volume}{86}},
  \bibinfo{pages}{1881} (\bibinfo{year}{2001}).

\bibitem[{\citenamefont{Misguich et~al.}(2002)\citenamefont{Misguich, Serban,
  and Pasquier}}]{misguich}
\bibinfo{author}{\bibfnamefont{G.}~\bibnamefont{Misguich}},
  \bibinfo{author}{\bibfnamefont{D.}~\bibnamefont{Serban}}, \bibnamefont{and}
  \bibinfo{author}{\bibfnamefont{V.}~\bibnamefont{Pasquier}},
  \bibinfo{journal}{Phys. Rev. Lett.} \textbf{\bibinfo{volume}{89}},
  \bibinfo{pages}{137202} (\bibinfo{year}{2002}).

\bibitem[{\citenamefont{Sutherland}(1988)}]{bsutherland}
\bibinfo{author}{\bibfnamefont{B.}~\bibnamefont{Sutherland}},
  \bibinfo{journal}{Phys. Rev. B} \textbf{\bibinfo{volume}{37}},
  \bibinfo{pages}{3786} (\bibinfo{year}{1988}).

\bibitem[{\citenamefont{Poilblanc, Mambrini, Schwandt}(2010)}]{poilblanc}
\bibinfo{author}{\bibfnamefont{D.} \bibnamefont{Poilblanc}},
  \bibinfo{author}{\bibfnamefont{M.} \bibnamefont{Mambrini}}, 
  \bibinfo{author}{\bibfnamefont{D.} \bibnamefont{Schwandt}},
  \bibinfo{journal}{Phys. Rev. B} \textbf{\bibinfo{volume}{81}},
  \bibinfo{pages}{180402(R)} (\bibinfo{year}{2010}).

\bibitem[{\citenamefont{Schwandt, Mambrini, Poilblanc}(2010)}]{schwandt}
\bibinfo{author}{\bibfnamefont{D.} \bibnamefont{Schwandt}},
  \bibinfo{author}{\bibfnamefont{M.} \bibnamefont{Mambrini}}, 
  \bibinfo{author}{\bibfnamefont{D.} \bibnamefont{Poilblanc}},
  \bibinfo{journal}{Phys. Rev. B} \textbf{\bibinfo{volume}{81}},
  \bibinfo{pages}{214413} (\bibinfo{year}{2010}).

\bibitem[{\citenamefont{Elser}(1989)}]{elser}
\bibinfo{author}{\bibfnamefont{V.}~\bibnamefont{Elser}},
  \bibinfo{journal}{Phys. Rev. Lett.} \textbf{\bibinfo{volume}{62}},
  \bibinfo{pages}{2405} (\bibinfo{year}{1989}).

\bibitem[{\citenamefont{Nussinov et~al.}(2007)\citenamefont{Nussinov, Batista,
  Normand, and Trugman}}]{nussinov}
\bibinfo{author}{\bibfnamefont{Z.}~\bibnamefont{Nussinov}},
  \bibinfo{author}{\bibfnamefont{C.~D.} \bibnamefont{Batista}},
  \bibinfo{author}{\bibfnamefont{B.}~\bibnamefont{Normand}}, \bibnamefont{and}
  \bibinfo{author}{\bibfnamefont{S.~A.} \bibnamefont{Trugman}},
  \bibinfo{journal}{Phys. Rev. B} \textbf{\bibinfo{volume}{75}},
  \bibinfo{pages}{094411} (\bibinfo{year}{2007}).

\bibitem[{\citenamefont{Fujimoto}(2005)}]{fujimoto}
\bibinfo{author}{\bibfnamefont{S.}~\bibnamefont{Fujimoto}},
  \bibinfo{journal}{Phys. Rev. B} \textbf{\bibinfo{volume}{72}},
  \bibinfo{pages}{024429} (\bibinfo{year}{2005}).

\bibitem[{\citenamefont{a_seidel09}(2009)}]{a_seidel09}
\bibinfo{author}{\bibfnamefont{A.}~\bibnamefont{Seidel}},
  \bibinfo{journal}{Phys. Rev. B} \textbf{\bibinfo{volume}{80}},
  \bibinfo{pages}{165131} (\bibinfo{year}{2009}).

\bibitem[{\citenamefont{Cano and Fendley}(2009)}]{cano}
\bibinfo{author}{\bibfnamefont{J.}~\bibnamefont{Cano}}, \bibnamefont{and}
  \bibinfo{author}{\bibfnamefont{P.}~\bibnamefont{Fendley}},
  \bibinfo{journal}{Phys. Rev. Lett} \textbf{\bibinfo{volume}{105}},
  \bibinfo{pages}{067205} (\bibinfo{year}{2010}).

\bibitem[{\citenamefont{Yao and Lee}(2010)}]{yao}
\bibinfo{author}{\bibfnamefont{H.}~\bibnamefont{Yao}} \bibnamefont{and}
  \bibinfo{author}{\bibfnamefont{D.-H.} \bibnamefont{Lee}},
  \bibinfo{journal}{arXiv:1010.3724}  (\bibinfo{year}{2010}).

\bibitem[{\citenamefont{Raman et~al.}(2005)\citenamefont{Raman, Moessner, and
  Sondhi}}]{decorate}
\bibinfo{author}{\bibfnamefont{K.~S.} \bibnamefont{Raman}},
  \bibinfo{author}{\bibfnamefont{R.}~\bibnamefont{Moessner}}, \bibnamefont{and}
  \bibinfo{author}{\bibfnamefont{S.~L.} \bibnamefont{Sondhi}},
  \bibinfo{journal}{Phys. Rev. B} \textbf{\bibinfo{volume}{72}},
  \bibinfo{pages}{064413} (\bibinfo{year}{2005}).


\bibitem[{\citenamefont{Flocke et~al.}(1998)\citenamefont{Flocke, Schmalz, and
  Klein}}]{klein}
\bibinfo{author}{\bibfnamefont{N.} \bibnamefont{Flocke}},
  \bibinfo{author}{\bibfnamefont{T.~G.}~\bibnamefont{Schmalz}}, \bibnamefont{and}
  \bibinfo{author}{\bibfnamefont{D.~J.} \bibnamefont{Klein}},
  \bibinfo{journal}{J. Chem. Phys.} \textbf{\bibinfo{volume}{109}},
  \bibinfo{pages}{873} (\bibinfo{year}{1998}).

\bibitem[{\citenamefont{Noorbakhsh et~al.}(2007)\citenamefont{Noorbakhsh, Shahbazi,
  Jafari, and Baskaran}}]{noorbakhsh}
\bibinfo{author}{\bibfnamefont{Z.}~\bibnamefont{Noorbakhsh}},
  \bibinfo{author}{\bibfnamefont{F.}~\bibnamefont{Shahbazi}},
  \bibinfo{author}{\bibfnamefont{S.~A.}~\bibnamefont{Jafari}}, \bibnamefont{and}
  \bibinfo{author}{\bibfnamefont{G.} \bibnamefont{Baskaran}},
  \bibinfo{journal}{Journal of the Physical Society of Japan} \textbf{\bibinfo{volume}{78}},
  \bibinfo{pages}{054701} (\bibinfo{year}{2009}).

\bibitem[{\citenamefont{Mambrini and Mila}(2000)}]{mambrini}
\bibinfo{author}{\bibfnamefont{M.}~\bibnamefont{Mambrini}} \bibnamefont{and}
  \bibinfo{author}{\bibfnamefont{F.}~\bibnamefont{Mila}},
  \bibinfo{journal}{Eur. Phys. J. B} \textbf{\bibinfo{volume}{17}},
  \bibinfo{pages}{651} (\bibinfo{year}{2000}).

\bibitem[{\citenamefont{Misguich1 et~al.}(2002)\citenamefont{Misguich1,
  Lhuillier, Mambrini, and Sindzingre}}]{ML2}
\bibinfo{author}{\bibfnamefont{G.}~\bibnamefont{Misguich}},
  \bibinfo{author}{\bibfnamefont{C.}~\bibnamefont{Lhuillier}},
  \bibinfo{author}{\bibfnamefont{M.}~\bibnamefont{Mambrini}}, \bibnamefont{and}
  \bibinfo{author}{\bibfnamefont{P.}~\bibnamefont{Sindzingre}},
  \bibinfo{journal}{Eur. Phys. J. B} \textbf{\bibinfo{volume}{26}},
  \bibinfo{pages}{167} (\bibinfo{year}{2002}).

\bibitem[{\citenamefont{Chayes et~al.}(1989)\citenamefont{Chayes, Chayes, and
  Kivelson}}]{kcc}
\bibinfo{author}{\bibfnamefont{J.~T.} \bibnamefont{Chayes}},
  \bibinfo{author}{\bibfnamefont{L.}~\bibnamefont{Chayes}}, \bibnamefont{and}
  \bibinfo{author}{\bibfnamefont{S.~A.} \bibnamefont{Kivelson}},
  \bibinfo{journal}{Commun. Math. Phys.} \textbf{\bibinfo{volume}{123}},
  \bibinfo{pages}{53} (\bibinfo{year}{1989}).

\bibitem[{\citenamefont{Misguich and Lhuillier}(2005)}]{ML}
\bibinfo{author}{\bibfnamefont{G.}~\bibnamefont{Misguich}} \bibnamefont{and}
  \bibinfo{author}{\bibfnamefont{C.}~\bibnamefont{Lhuillier}},
  \emph{\bibinfo{title}{Frustrated Spin Systems, {\em edited by H.T. Diep}}}
  (\bibinfo{publisher}{World Scientific}, \bibinfo{address}{Singapore},
  \bibinfo{year}{2005}), \urlprefix\url{http://arxiv.org/abs/cond-mat/0310405}.

\bibitem[{\citenamefont{Moessner and Sondhi}(2001)}]{raman1}
\bibinfo{author}{\bibfnamefont{R.}~\bibnamefont{Moessner}} \bibnamefont{and}
  \bibinfo{author}{\bibfnamefont{S.~L.} \bibnamefont{Sondhi}},
  \bibinfo{journal}{Phys. Rev. B} \textbf{\bibinfo{volume}{63}},
  \bibinfo{pages}{224401} (\bibinfo{year}{2001}).

\bibitem[{\citenamefont{Ziff and Scullard}(2006)}]{ziff}
\bibinfo{author}{\bibfnamefont{R.~M.}~\bibnamefont{Ziff}} \bibnamefont{and}
  \bibinfo{author}{\bibfnamefont{C.~R.} \bibnamefont{Scullard}},
  \bibinfo{journal}{J. Phys. A: Math. Gen.} \textbf{\bibinfo{volume}{39}},
  \bibinfo{pages}{15083} (\bibinfo{year}{2006}).


\bibitem[{not({\natexlab{a}})}]{note1}
\bibinfo{note}{{{While here we will usually consider the case where a 
 single type of cell suffices, one may in some cases want to identify several types 
 of cells that satisfy the local linear independence property, such that the lattice 
 can be conveniently covered by this family of cells.
  } }}


\bibitem[{not({\natexlab{a}})}]{note2}
\bibinfo{note}{{{In general, one should exclude {\em by definition} any link between two 
 boundary sites of $\cC$ from the topology of $\cC$. This is not an issue for most cells considered here, 
 except for the martini-B and the archimedean-B lattice, see \Fig{martini_middleright} (a),(c), \Fig{arch} (b),(d) . It would also be relevant, e.g., to cells 
 of the triangular lattice.
  } }}

\bibitem[{Lin(2008)}]{LinBox}
\emph{\bibinfo{title}{{LinBox -- Exact Linear Algebra over the Integers and
  Finite Rings, Version 1.1.6}}}, \bibinfo{organization}{The LinBox~Group}
  (\bibinfo{year}{2008}), \urlprefix\url{{http://linalg.org}}.

 
 \bibitem[{not({\natexlab{a}})}]{note3}
\bibinfo{note}{{{Strictly speaking, since we define a lattice
 only through its vertices and edges, while faces play no role,
 we can equally well regard the $C_{60}$ lattice as having
 the topology of a sphere, or, through its Schlegel diagram, of a planar graph.
 This does not affect the general validity of this statement.}
   } }
\bibitem[{\citenamefont{Rumer}(1932)}]{rumer}
\bibinfo{author}{\bibfnamefont{G.}~\bibnamefont{Rumer}},
  \bibinfo{journal}{{G\"ottinger. Nachr.}} p. \bibinfo{pages}{377}
  (\bibinfo{year}{1932}).

\bibitem[{\citenamefont{Pauling}(1933)}]{pauling}
\bibinfo{author}{\bibfnamefont{L.}~\bibnamefont{Pauling}}, \bibinfo{journal}{J.
  Chem. Phys.} \textbf{\bibinfo{volume}{1}}, \bibinfo{pages}{280}
  (\bibinfo{year}{1933}).

\bibitem[{\citenamefont{Mazumdar and Soos}(1979)}]{soos1}
\bibinfo{author}{\bibfnamefont{S.}~\bibnamefont{Mazumdar}} \bibnamefont{and}
  \bibinfo{author}{\bibfnamefont{Z.~G.} \bibnamefont{Soos}},
  \bibinfo{journal}{Synthetic Metals} \textbf{\bibinfo{volume}{1}},
  \bibinfo{pages}{77} (\bibinfo{year}{1979}).

\bibitem[{\citenamefont{Ramasesha and Soos}(1984)}]{soos2}
\bibinfo{author}{\bibfnamefont{S.}~\bibnamefont{Ramasesha}} \bibnamefont{and}
  \bibinfo{author}{\bibfnamefont{Z.~G.} \bibnamefont{Soos}},
  \bibinfo{journal}{Int. J. Quantum Chem.} \textbf{\bibinfo{volume}{25}},
  \bibinfo{pages}{1003} (\bibinfo{year}{1984}).

\bibitem[{\citenamefont{Okamoto et~al.}(2007)\citenamefont{Okamoto, Nohara, Aruga-Katori, and
  Takagi}}]{hyperkg_experiment}
\bibinfo{author}{\bibfnamefont{Y.}~\bibnamefont{Okamoto}},
  \bibinfo{author}{\bibfnamefont{M.}~\bibnamefont{Nohara}},
  \bibinfo{author}{\bibfnamefont{H.}~\bibnamefont{Aruga-Katori}}, \bibnamefont{and}
  \bibinfo{author}{\bibfnamefont{H.} \bibnamefont{Takagi}},
  \bibinfo{journal}{Phys. Rev. Lett} \textbf{\bibinfo{volume}{99}},
  \bibinfo{pages}{137207} (\bibinfo{year}{2007}).

\bibitem[{\citenamefont{Chen and Balents}(2009)}]{hyper1}
\bibinfo{author}{\bibfnamefont{G.}~\bibnamefont{Chen}}, \bibnamefont{and}
  \bibinfo{author}{\bibfnamefont{L.}~\bibnamefont{Balents}},
  \bibinfo{journal}{Phys. Rev. B} \textbf{\bibinfo{volume}{78}},
  \bibinfo{pages}{094403} (\bibinfo{year}{2008}).

\bibitem[{\citenamefont{Lawler et~al.}(2008)\citenamefont{Lawler, Kee, Kim, and
  Vishwanath}}]{hyper2}
\bibinfo{author}{\bibfnamefont{M.~J.}~\bibnamefont{Lawler}},
  \bibinfo{author}{\bibfnamefont{H.~-Y.}~\bibnamefont{Kee}},
  \bibinfo{author}{\bibfnamefont{Y.~B.}~\bibnamefont{Kim}}, \bibnamefont{and}
  \bibinfo{author}{\bibfnamefont{A.} \bibnamefont{Vishwanath}},
  \bibinfo{journal}{Phys. Rev. Lett} \textbf{\bibinfo{volume}{100}},
  \bibinfo{pages}{227201} (\bibinfo{year}{2008}).

\bibitem[{\citenamefont{Zhou et~al.}(2008)\citenamefont{Zhou, Lee, Ng, and
  Zhang}}]{hyper3}
\bibinfo{author}{\bibfnamefont{Y.}~\bibnamefont{Zhou}},
  \bibinfo{author}{\bibfnamefont{P.~A.}~\bibnamefont{Lee}},
  \bibinfo{author}{\bibfnamefont{T.~-K.}~\bibnamefont{Ng}}, \bibnamefont{and}
  \bibinfo{author}{\bibfnamefont{F.~-C.} \bibnamefont{Zhang}},
  \bibinfo{journal}{Phys. Rev. Lett} \textbf{\bibinfo{volume}{101}},
  \bibinfo{pages}{197201} (\bibinfo{year}{2008}).

\bibitem[{\citenamefont{Lawler2 et~al.}(2008)\citenamefont{Lawler, Paramekanti, Kim, and
  Balents}}]{hyper4}
\bibinfo{author}{\bibfnamefont{M.~J.}~\bibnamefont{Lawler}},
  \bibinfo{author}{\bibfnamefont{A.}~\bibnamefont{Paramekanti}},
  \bibinfo{author}{\bibfnamefont{Y.~B.}~\bibnamefont{Kim}}, \bibnamefont{and}
  \bibinfo{author}{\bibfnamefont{L.} \bibnamefont{Balents}},
  \bibinfo{journal}{Phys. Rev. Lett} \textbf{\bibinfo{volume}{101}},
  \bibinfo{pages}{197202} (\bibinfo{year}{2008}).

\bibitem[{\citenamefont{Bergholtz et~al.}(2010)\citenamefont{Bergholtz, Laeuchli, and
  Moessner}}]{hyper5}
\bibinfo{author}{\bibfnamefont{E.~J.}~\bibnamefont{Bergholtz}},
  \bibinfo{author}{\bibfnamefont{A.~M.}~\bibnamefont{Laeuchli}}, \bibnamefont{and}
  \bibinfo{author}{\bibfnamefont{R.} \bibnamefont{Moessner}},
  \bibinfo{journal}{Phys. Rev. Lett} \textbf{\bibinfo{volume}{105}},
  \bibinfo{pages}{237202} (\bibinfo{year}{2010}).



\end{thebibliography}
\end{document}